\documentclass[nohyper,letterpaper,11pt,notoc]{JHEP3}
\usepackage{graphicx}
\usepackage{amssymb}
\usepackage{amsmath}
\usepackage{amsfonts}
\usepackage{latexsym}
\usepackage[all]{xy}
\usepackage{slashed}
\preprint{}

\newcommand{\keV}{~\mathrm{keV}}
\newcommand{\MeV}{~\mathrm{MeV}}
\newcommand{\GeV}{~\mathrm{GeV}}
\newcommand{\TeV}{~\mathrm{TeV}}
\newcommand{\cm}{~\mathrm{cm}}
\newcommand{\exE}{\delta}
\newcommand{\mN}{m_{\scriptscriptstyle N}}
\newcommand{\mX}{m_{\scriptscriptstyle \chi}}
\newcommand{\rhoX}{\rho_{\scriptscriptstyle \chi}}
\newcommand{\mXp}{m_{\scriptscriptstyle \chi^\prime}}

\newcommand{\vsun}{v_{\odot}}
\newcommand{\eqtau}{\tau_{\scriptscriptstyle {\rm eq}}}

\title{Capture of Inelastic Dark Matter in the Sun} 
\author{Shmuel Nussinov$^{(a,b)}$, Lian-Tao Wang$^{(c)}$ and Itay
  Yavin$^{(c)}$\\ \it{(a) School of Physics and Astronomy, Tel Aviv University, Tel Aviv, Israel} \\ \it{(b) Schmid College of Science, Department of Physics, Chapman University, Orange, CA, 92855}\\\it{(c) Department of Physics, Princeton University, Princeton, NJ 08544}} 

\abstract{We consider the capture of dark matter in the Sun by inelastic scattering against nuclei as in the inelastic dark matter scenario. We show that, assuming a WIMP-nucleon cross-section of $\sigma_n = 10^{-40}\cm^2$ the resulting capture rate and density are sufficiently high so that current bounds on the muon neutrino flux from the Sun rule out any appreciable annihilation branching ratio of WIMPs into $W^+W^-$, $Z^0Z^0$, $\tau^+\tau^-$, $t\bar{t}$ and neutrinos.  Slightly weaker bounds are also available for annihilations into $b\bar{b}$ and $c\bar{c}$. Annihilations into lighter particles that may produce neutrinos, such as $\mu^+\mu^-$, pions and kaons are unconstrained since those stop in the Sun before decaying. Interestingly enough, this is consistent with some recent proposals motivated by the PAMELA results for the annihilation of WIMPs into light bosons which subsequently decay predominantly into light leptons and pions. } 

\begin{document}
\section{Introduction}

In the past twenty or so years, much effort has been devoted to probing galactic dark matter, (DM) beyond its gravitational effects, in an attempt to uncover what consitutes DM and how it interacts. Maybe the most promising candidate is a weakly interacting massive particle (WIMP). Direct-detection searches look for WIMP-nucleus scattering in underground detectors and carry with them the hope of having laboratory control over DM. So far, most experiments have reported null results and placed very strong bounds on the elastic cross-section for WIMP scattering against nuclear matter. Notwithstanding these results, the DAMA collaboration announced an 8.3 $\sigma$ discovery in the annual modulations of nuclear recoil rate~\cite{Bernabei:2008yi}. Unfortunately, if attributed to WIMP elastic scattering, such event rates are excluded by more than two orders of magnitude by the other experiments. An intriguing resolution of this controversy is the proposal of ``inelastic dark matter'' (iDM) laid forth in Ref. \cite{TuckerSmith:2001hy}. If DM scattering off of nuclear matter is inelastic, with a transition to an excited state of DM roughly $100\keV$ above the ground state, then the incosistency between the different experimental results can be settled~\cite{Chang:2008gd}. 

Alongside the direct detection efforts, there are also indirect searches looking for energetic neutrinos from WIMP annihilations in the Sun or the Earth or other energetic fluxes (photons, positrons, and antiprotons) from annihilations in the intragalactic medium. Very early on, in a seminal paper, Press and Spergel~\cite{Press:1985ug} considered the possibility of WIMPs' capture by the Sun and their computation was later refined and corrected by Gould~\cite{Gould:1987ir}. In their work, and all subsequent work on the subject, capture was assumed to proceed through elastic collisions of WIMPs with matter. In this paper we consider the iDM scenario where capture proceeds via inelastic scattering. We compute the associated capture rate and discuss detection prospects in neutrino telescopes.

It is important to realize that the computation of the neutrino flux, starting from  WIMPs' capture and ending with their annihilations, involves several steps which are model dependent. The problem nicely divides into a question of capture and a question of annihilation. These are logically disjoint and call for a modular approach which we carefully review in section~\ref{sec:steps}. Throughout, we strive to maintain a model independent approach and clearly state where and which assumptions are made.

In section~\ref{sec:recent} we make a short detour and consider the relation of the results presented in this paper to some of the recent developments associated with DM observations and model building. Section~\ref{sec:steps} is devoted to the formulation of the problem and the modular approach we adopt in resolving it. In section~\ref{sec:results} we present and discuss the results and section~\ref{sec:conclusions} contains our conclusions. 

Finally, we note that in their original work on iDM, Smith and Weiner already considered the possibility of capture of WIMPs in the Sun. Our quantitative conclusions differ from theirs on three accounts. First, we include form-factor effects which are extremely important for iron. In the elastic case, these effects are so significant as to dethrone iron from its place as the prime capturer of WIMPs. Second, we take into account the case where elastic scattering is altogether absent and show that the resulting density is nevertheless sufficiently high to saturate the annihilation rate in the Sun. This ensures maximal signal and closes a loophole whereby iDM models with little or no elastic component can escape the bounds presented below. Third, the oscillations of the different neutrino flavors into each other are now well established. This closes another loophole whereby sneutrino type WIMPs annihilate into electron-neutrino only. Previously, the possibility existed for these neutrinos to keep their flavor and therefore result in no upward going muons in Earth-bound detectors.  

\section{Recent Developments}
\label{sec:recent}

While this work deals specifically with the problem of inelastic capture of WIMPs by the Sun, we momentarily diverge off this route to discuss its relation to some of the recent developments related to DM. This discussion is useful on its own right because it serves to illustrate the different ingredients that enter the computation and expounded upon in later sections. 

Motivated by the positron excess seen by PAMELA~\cite{Adriani:2008zr}, a new class of DM models was suggested in Ref.~\cite{ArkaniHamed:2008qn}, employing a light mediator that both explains the large cross-section and the absence of any excess in the anti-proton flux~\cite{Adriani:2008zq}. A very reasonable candidate for this mediator is the gauge-boson of an additional (non) abelian gauge group under which DM is charged. These models have the added benefit that the small splitting ($\sim 100\keV$) and inelastic coupling associated with iDM are naturally generated\footnote{The idea was first suggested in Ref.~\cite{ArkaniHamed:2008qn} for the non-abelian models and expanded upon in Ref.~\cite{Baumgart:2009tn}. The existence of this splitting in abelian models was demonstrated in Refs.~\cite{Katz:2009qq,Cheung:2009qd}.}. These models form an organic whole, capable of explaining many of the astrophysical anomalies reported in the recent years, and contain several distinct phenomenological signatures important for this work,

\begin{enumerate}
\item Sommerfeld enhancement of the annihilations of slow WIMPs via a relatively light boson helps to reconcile the large rate needed to explain the PAMELA positron excess with the cross-section deduced from the thermal relic abundance of DM.

\item WIMPs annihilate predominantly into a pair of these bosons which in turn are weakly mixed with the SM. Their decays into $\mu^+\mu^-$ or $e^+ e^-$ are responsible for the excesses seen in PAMELA, and possibly ATIC, whereas decays into anti-protons are kinematically suppressed~\cite{Cholis:2008vb}.

\item The features seen in PAMELA (and possibly ATIC) require a fairly heavy WIMP, anywhere between $100\GeV - \TeV$. 

\item DM excited states are naturally present in the spectrum with $\exE =\mXp-\mX \sim \mathcal{O}( 100~\mathrm{keV}-\mathrm{MeV})$ which can realize the iDM scenario.
\end{enumerate}

Nevertheless each of the above elements can independently arise in specific constructions. Hence, from a phenomenological point of view we should consider how each of these elements seperately impacts the prospects of indirect detection of WIMPs capture and annihilation in the Sun in neutrino telescopes experiments.  

Indeed, such a signal may be more robust than any of the other indirect probes. Anomalies in electromagnetic signals (PAMELA, ATIC) may be due to special astronomical mechanisms rather than WIMP annihilations. On the other hand a putative signal of energetic muon-neutrinos coming from the Sun during the (southern) winter or straight up from Earth's center may have no other reasonable explanation other than WIMP annihilations in the Sun or Earth\footnote{To put the issue of particle versus astrophysical explanation in perspective, recall the hotly debated
"Solar Neutrino Anomaly". Some of its  experimental evidences were suspected and astrophysical explanations suggested. The late John Bahcall correctly argued that while some experiments suggesting neutrino oscilations for certain neutrino energies may have astrophysical explanations the sum total of all evidences cannot \cite{Bludman:1993tk}. The present "Unified DM models" motivated by several anomalies in the electromagnetic spectrum from microwaves to multi-GeV Gammas may survive in some form or another. Note that also the solar case involved a more robust neutrino signal.}.

The first ingredient involving the Sommerfeld enhanced annihilation cross-section is in fact a boon since it helps to gaurantee equilibrium as discussed in section \ref{sec:results}. In particular it can greatly enhance the rate of neutrinos from the Earth as discussed in Ref. \cite{Delaunay:2008pc}.

The second ingredient concerning the annihilation channels has the most dramatic effect as far as neutrino telescopes are concerned. If the WIMPs dominantly annihilate into light bosons which consequently decay into muons and electrons, then \textit{no energetic neutrinos} will be observed. Electrons do not yield neutrinos, and the neutrinos from the muons' decay will be too soft since the muons first stop in the Sun and only then decay, yielding a Michel spectrum $\sim~30\MeV$ neutrinos~\cite{Ritz:1987mh,Crotty:2002mv}. So this class of models, in their purest form, lead to no observable consequences for neutrino telescopes\footnote{This is also true for models where WIMPs dominantly annihilate into photons, gluons, and kaons, all of which result in little if any energetic neutrinos.}. In what follows, we will instead remain agnostic about the annihilation channels open for WIMPs and quote results based on branching fractions. 

The third ingredient pertaining to the mass of the WIMP is relevant for the capture rate (and only mildly to the equilibrium condition). The capture rate is reduced by roughly $\sim \mX^{-2}$. One power of the mass is coming from the obvious inverse dependence of the WIMPs number density on their mass, once the local energy density is fixed, $\rhoX = 0.3\GeV\cm^{-3}$. A second power of the WIMPs mass is present because it becomes increasingly difficult to transfer enough momentum and gravitationally capture heavier WIMPs. The heaviest target in the Sun is iron so the capture rate for WIMPs with $\mX \gg 52\GeV$ is strongly suppressed. The resonant enhancement in the Earth discussed by Gould \cite{Gould:1987ir} is altogether absent. 

This reduction in capture rate is, however, offset by the increased neutrino energy. First, the conversion probability of neutrinos into muons in the rock or ice below the detector grows as $\sigma(\nu_{\mu} + N \rightarrow \mu + X) \sim E_{\nu}$. Second, the range of muons grows with their energy as well $\sim E_{\mu}$. This offset does not occur for annihilations of WIMPs in the Earth. The density of accumulated WIMPs in the Earth does not build to be ``optically thick'' to neutrinos and hence the signal is proportional to the the square of the capture rate and massive WIMPS yield a weaker signal. These considerations omit two other factors: i) The Sun becomes optically thick for a neutrino of energy $E_{\nu} > 1/2\TeV$ produced at the solar core, reducing the solar signal; ii) More energetic muons point more precisely in the direction of the parent neutrinos. The $10^{-3}$ radian precision required to resolve the solar core is unavilable. However better pointing improves the Earth signal since the annihilations occur at the center and this can be well tested by the long vertical strings of photomultiplyers in IceCube.

The fourth ingredient we comment on is the most relevant for the analysis in this paper, namely the inelastic transition. As clearly explained in \cite{TuckerSmith:2001hy}, an inelastic transition of $\sim100\keV$ results in a preference towards heavier targets and is essentially how the absence of nuclear recoils in CDMS (germanium) can be reconciled with the DAMA (iodine) results. Since iron is lighter than germanium the rate for capture in the Earth is tiny at best (nickel is actually a little heavier than iron, but its abundance is two orders of magnitude smaller). The situation would have been just as bad in the Sun if it was not for the kinetic energy gained by the WIMP as it falls in the Sun's gravitational well. The escape velocity in the Sun ranges in $v_{esc} \sim 600 - 1300~{\rm km/s}$ and provides sufficient energy to overcome the excitation barrier of $\sim 100\keV$. This entire paper rests on this simple observation, one that was made already in the original work of Smith and Weiner \cite{TuckerSmith:2001hy}.

Finally, we close this section with a brief survey of existing iDM models and their relevancy to this paper. Rather than concentrating on the DM identity, a more useful categorization is obtained by focusing on the mediator of the inelastic transition. One possible mediator is the SM $Z^0$ as happens for example in the case of sneutrino DM~\cite{TuckerSmith:2001hy} (see \cite{Cui:2009xq, Alves:2009nf} for many other interesting possibilities). If kinematically allowed, DM can then annihilate to a pair of $Z^0$'s, which as we show below is strongly constrained. Similar comments would apply for a heavier $Z^\prime$, however, in this case the annihilation may be kinematically suppressed. Mediation through a scalar (higgs or other) is somewhat less constrained since DM annihilation into this scalar may produce only bottom quarks as final products. As already mentioned above, the possibility of a new light mediator can avoid the bounds entirely. DM annihilation in the Sun into any light entity ($\lesssim\GeV$) would result in no observable neutrinos. Other variants are certainly possible, and we hope that the modular approach of this paper will allow for direct comparison with any other future models of iDM.

\section{Methodology}
\label{sec:steps}
\subsection{Approach to Equilibrium}

The time evolution of the WIMP population in the Sun is controled by (we follow the notation of Ref. \cite{Jungman:1995df}),
\begin{equation}
\label{eqn:popDE}
\dot{N} = C - C_{A} N^2
\end{equation}
where $C$ is the capture rate of WIMPs in the Sun and $C_{A}$ is related to the annihilation rate as $\Gamma_{A} =\frac{1}{2} C_A N^2$. The evaluation of the capture rate, $C$, will be presented below. In this subsection we concentrate on the estimation of $C_A$ and the modification necessary for inelastic scattering. Its significance manifests itself by solving Eq. (\ref{eqn:popDE}) for the annihilation rate at a given time,
\begin{equation}
\label{eqn:AnnRate}
\Gamma_A = \frac{1}{2} C \tanh^2\left(t/\eqtau \right)
\end{equation}
where $\eqtau = 1/\sqrt{C C_A}$ is the time scale required to reach equilibrium between capture and annihilation. When $\eqtau$ becomes comparable or larger than the age of the solar system $t_{\odot} \simeq 1.5\times 10^{17}~s$, the system has not yet reached equilibrium and the annihilation rate is strongly suppressed,
\begin{equation}
\label{eqn:DimAnnRate}
\Gamma_A \approx \frac{1}{2} C \left(t/\eqtau \right)^2
\end{equation}
On the other hand, when $t_{\odot} \gg \eqtau$, the annihilation rate is saturated at $\Gamma_A = \frac{1}{2} C$. 

In the usual WIMP scenario with elastic scattering against nuclear matter, $C_A$ is essentially determined by the temperature of the Sun and the annihilation cross-section. This is based on the assumption that after being gravitationally captured, WIMPs will continue to lose energy with every subsequent collision and reach thermal equilibrium in the Sun's core. Even with elastic cross-sections as low as $\sigma_n = 10^{-43}\cm^2$ the number of collisions they undergo over the solar lifetime is $\sim t_{\odot} \langle \sigma_n v \rangle n_{\odot} = 3\times10^7$, which is sufficient to thermalize. In that case, their number density is given by \cite{Griest:1986yu},
\begin{equation}
\label{eqn:ThermNumDens}
n(r) = n_0 e^{-\mX \phi(r)/T} = n_0 e^{-r^2/r_{{\rm th}}^2}
\end{equation}
where $n_0$ is their number density in the center of the Sun, $\phi(r)$ is the gravitational potential with respect to the core, and $T$ is their temperature. The second equality holds if we assume a constant core density $\rho$ and define the thermal radius,
\begin{equation}
\label{eqn:thradius}
r_{{\rm th}} = \left(\frac{3 T}{2\pi \mX G \rho}\right)^{1/2} = 0.01\times R_{\odot} \left(\frac{T}{1.2\keV}\right)^{1/2} \left(\frac{100\GeV}{\mX}\right)^{1/2}
\end{equation}
where $R_{\odot}$ is the Sun's radius and $T = 1.2\keV$ is the Sun's core temperature. The annihilation rate per WIMP squared is given by,
\begin{equation}
C_A = \frac{\int d^3r~n(r)^2 ~\langle \sigma_A v \rangle }{\left(\int d^3r~ n(r) \right)^2 } = \frac{\langle\sigma_A v\rangle}{(2\pi)^{3/2} r_{th}^3}
\end{equation}
where $\sigma_A$ is the annihilation cross-section, and the last equality follows from Eq. (\ref{eqn:ThermNumDens}). The ratio of solar age to equilibrium time is then given by, 
\begin{equation}
\label{eqn:eqbratio}
\frac{t_{\odot}}{\eqtau} = 10^{3} \left(\frac{C}{10^{25}\sec^{-1}}\right)^{1/2} \left( \frac{\langle \sigma_A v \rangle}{ 3 \times 10^{-26}\cm^3\sec^{-1}}\right)^{1/2}\left(\frac{0.01\times R_\odot}{r_{th}}\right)^{3/2}
\end{equation}
which implies that equilibrium has been reached long ago and we can expect the full signal from the Sun.

The situation is more subtle in the iDM scenario. First, let us consider the case where a small elastic component exists in the scattering of WIMPs against nuclear targets. In this case, while capture proceeds through an inelastic transition (with a large cross-section), the subsequent thermalization of the WIMP against the nucleons in the Sun can be due to its suppressed ($\sigma_n < 10^{-43}\cm^2$) elastic scattering. In specific models of iDM, the elastic coupling is suppressed with respect to the inelastic one by $\sim 10^{-6}$ \footnote{This is usually achieved by starting with a Dirac fermion (or a complex scalar) of mass $\mX$ and adding a small Majorana mass $m$. That splits the fermion into its Majorana components and generates a mass splitting of $\mXp - \mX =\exE = m^2/\mX \approx 100\keV$. Any vector current coupeled to the fermion will result in a dominantly inelastic coupling between $\chi$ and $\chi^\prime$. A small elastic coupling of size $\exE/\mX \approx 10^{-6}-10^{-7}$ is also generated.}. That would result in an elastic cross-section (per nucleon) of $\sigma_n \simeq 10^{-52}\cm^2$ which is too small to bring the WIMPs into thermal equilibrium with the rest of the matter in the Sun. However, if we take the elastic cross-section to be $\sigma \lesssim 10^{-43}\cm^2$ (current bound from direct-detection), but $\sigma_n > 10^{-47}\cm^2$ the WIMP can undergo enough collisions to thermalize. In that case, the above estimates for the annihilation rate $C_A$ still apply. 

As was recently clarified and emphasized in Ref. \cite{Batell:2009vb}, even in the absence of any elastic coupling, the second term in the Born series yields an elastic transition. When the force mediator is very light $\lesssim 50\MeV$, the resulting elastic cross-section per nucleon is sufficiently large ($\sim 10^{-42}\cm^2$) to allow for thermalization. However, the cross-section drops rapidly as the mass of the mediator increases and becomes inefficient for a mass of $\sim \GeV$. 

In order to close any possible loophole in the argument we computed the final WIMPs' density in the pure inelastic case where no elastic scattering is allowed. In order to facilitate the computation,  we approximated the Sun's gravitational potential with an analytic form which allows for an exact solution of the oribts\footnote{We would like to extend our warm gratitude to D. Lynden-Bell for bringing this solution to our attention.}, as explicated in a beautiful paper by Henon \cite{Henon1,Henon2}. We describe the details of the computation in appendix \ref{app:density}. In Fig. \ref{fig:density} we depict the resulting density for a particular choice of parameters. The WIMPs undergo only few collisions before their kinetic enery anywhere in the orbit drops below the inelastic threshold. Therefore, it is difficult for them to shed off all their angular momentum which is the reason why the density vanishes as we approach the Sun's center. On the other hand, since most of the Sun's mass is concentrated in its core, most collisions actually happen in the inner radius ($<0.2 R_{\odot}$) and the resulting orbits are to a large extent contained inside the Sun as shown in Fig. \ref{fig:orbit}. As the WIMP gets heavier, it become harder for it to shed off kinetic energy in every collision, but since its initial kinetic energy is larger, it also undergoes more collisions. Due to the different approximations used in arriving at these results, they are probably only good to $20\%$ accuracy. However, considering that $\tanh(t_{\odot}/\eqtau)\approx 1$ for $t_{\odot}/\eqtau\gg 1$, these uncertainties do not propagate into the annihilation rate and muon yield discussed below. We conclude that even in the worst case scenario where no elastic scattering is allowed, the annihilation rate still reaches equiliberium and is saturated by half the capture rate\footnote{We have assumed here that after its first excitation collision, the WIMP quickly deexcite to the ground state so any subsequent collision is also endothermic. If this is not the case, and the deexcitation time is very long (as recently discussed in Ref. \cite{Finkbeiner:2009mi,Batell:2009vb}), the situation is even more favorable since the WIMP can now scatter against all the nuclei in the Sun, including hydorgen and helium.}.  
\vspace{10mm}
\begin{figure}[h]
\begin{center}
\includegraphics[scale=.3]{WIMPdensity.eps}
\includegraphics[scale=.55]{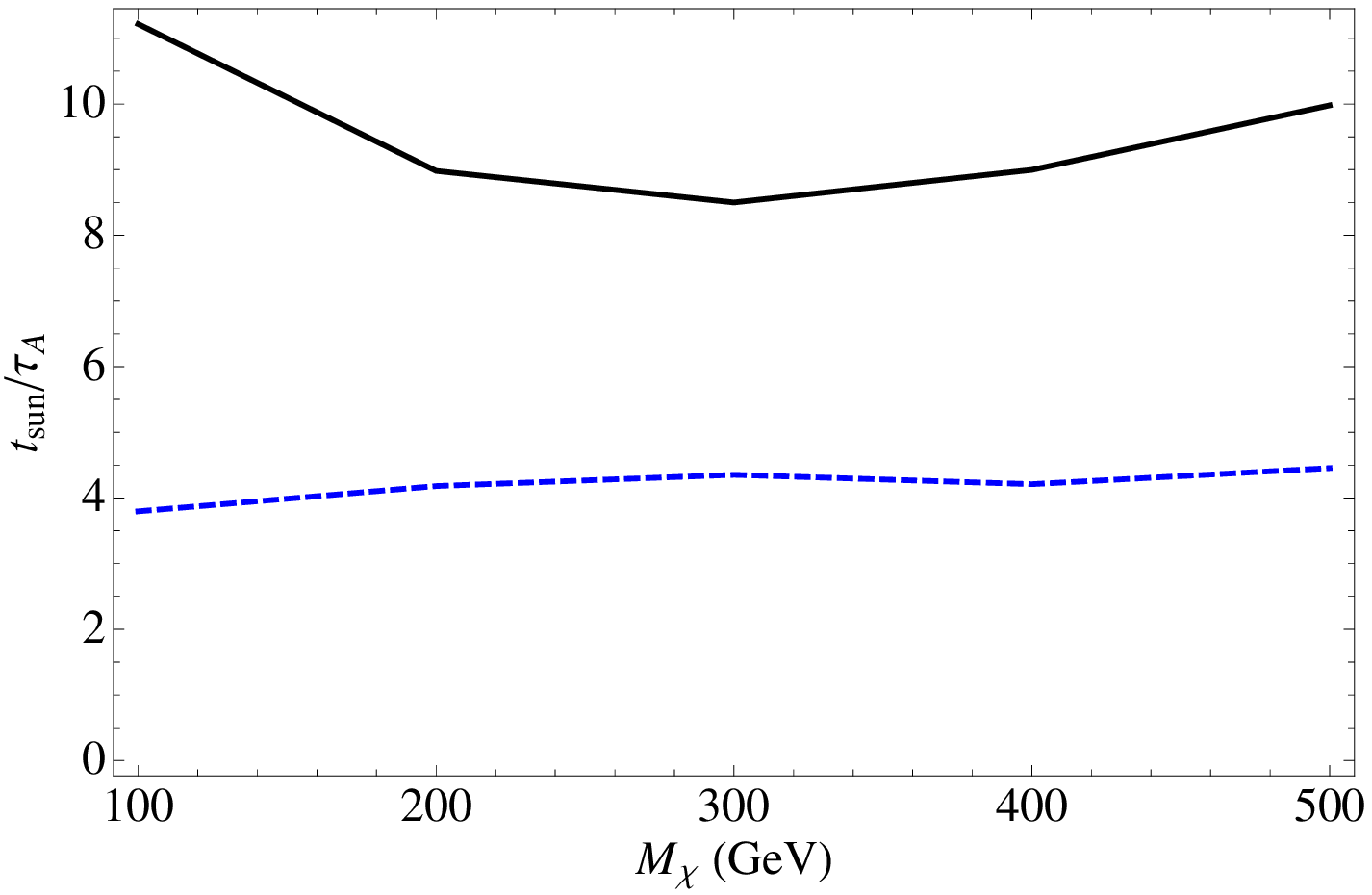}
\end{center}
\caption{On the left pane we show the WIMP's density (normalized to unity) against the distance from the Sun's center for $\mX = 100\GeV$ and $\exE = 100\keV$ (solid-black), $\exE = 150\keV$ (dashed-blue), and $\exE=200\keV$ (dotdashed-red). On the right pane we depict the ratio $t_{\odot}/\eqtau$ as a function of the WIMP's mass for $\exE=100\keV$ (black-solid) and $\exE=150\keV$ for $\langle\sigma_A v \rangle = 3\times 10^{-26}\cm^3\rm{~sec^{-1}}$ and $C = 10^{25}\rm{~sec^{-1}}$ as in Eq. (\ref{eqn:eqbratio}).}
\label{fig:density}
\end{figure}

\subsection{Capture Rate}

If equilibrium is reached then WIMPs' annihilation rate is saturated at half the capture rate $\Gamma_A = \frac{1}{2}C$. The formulae and procedure related to the computation of the inelastic capture rate, $C$, are presented in appendix \ref{app:caprate}. In here we discuss some of the qualitative points that arise and clarify the modifications involved. For concretness, we fix the inelastic energy threshold to be $\exE = 100 \keV$ whenever numerical values are used. 

\begin{figure}[h]
\begin{center}
\includegraphics[scale=.5]{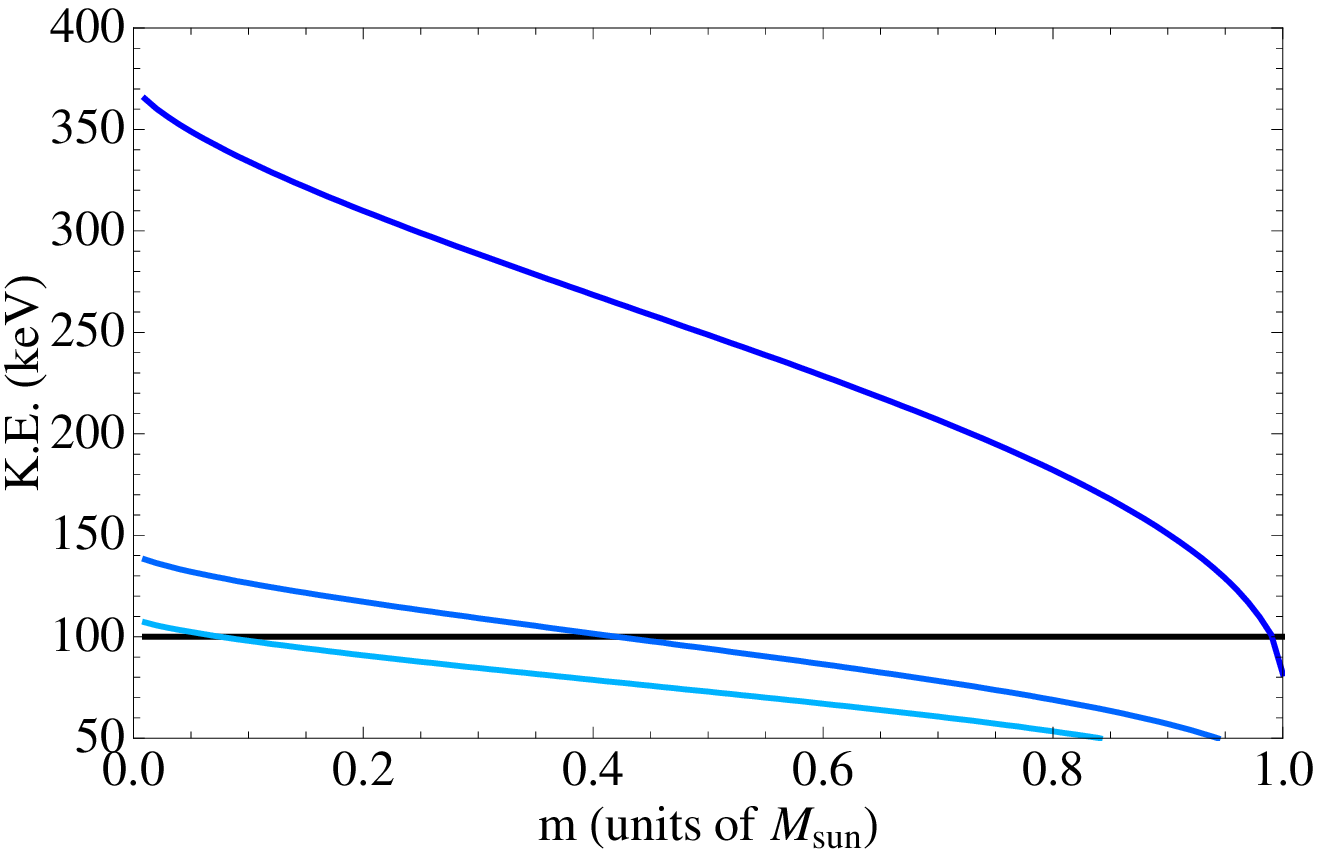}
\includegraphics[scale=.5]{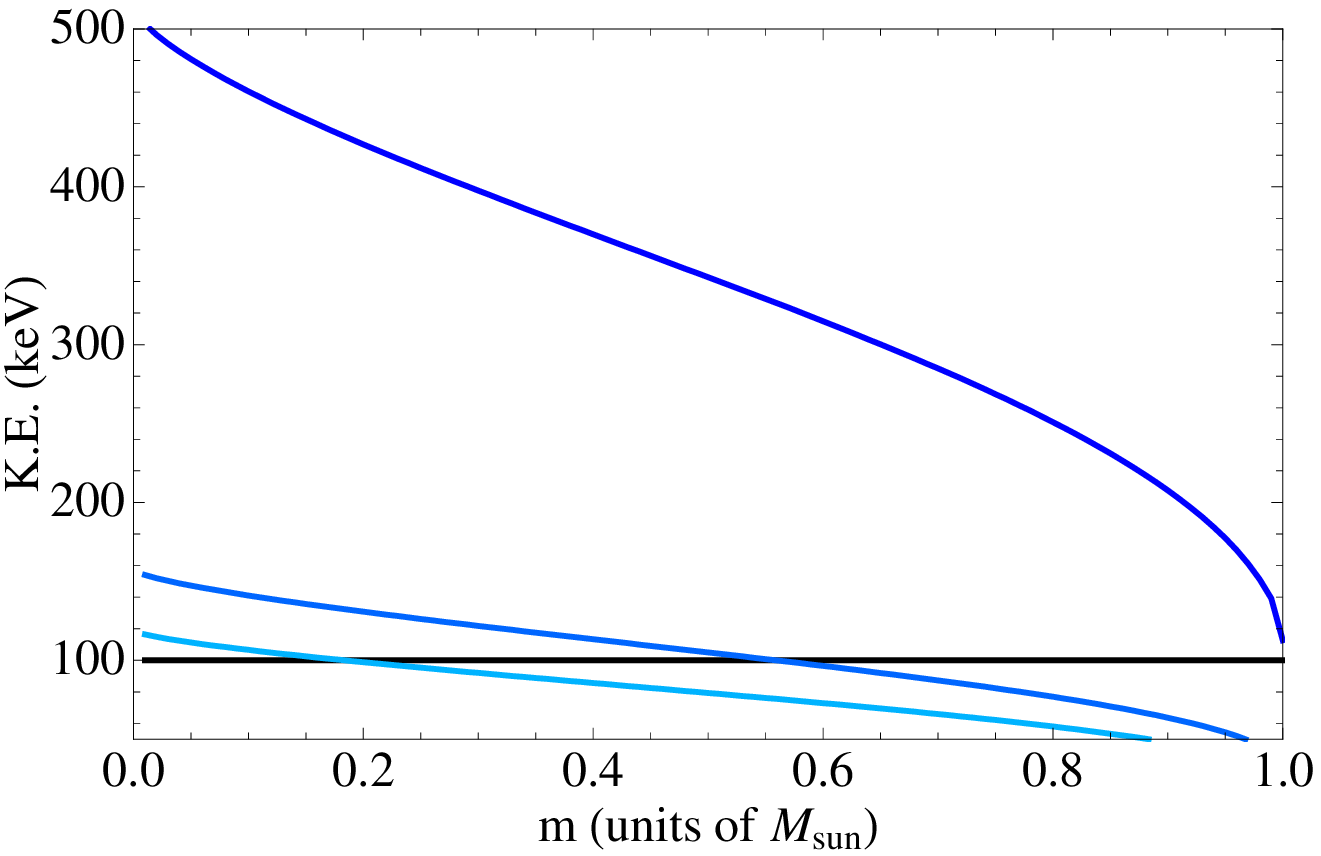}
\end{center}
\caption{The kinetic energy in the WIMP-nucleus center of mass frame as a function of the mass $m$ (in units of $M_{\odot}$) enclosed in a given shell. The velocity at infinity was taken to be $u = 220~{\rm km/s}$ as explained in the text. The curves correspond to scattering against iron, oxygen, and carbon from top to bottom. The WIMP mass was fixed at $100\GeV$ ($500\GeV$) on the left (right) pane. Carbon is the lightest element present in the Sun against which the WIMP can scatter inelastically.}
\label{fig:KEvsM}
\end{figure}

As the WIMPs fall into the Sun, they get accelerated in the gravitational well until they meet a nucleus against which they collide. Their velocity at that point is given by,
\begin{equation}
w(r)^2 = u^2 + v(r)^2
\end{equation}
where $u$ is their velocity at inifinity, determined by the velocity distribution in the halo and the Sun's motion through the galaxy, and $v(r)$ is the escape velocity from the Sun at radius$~r$. For a WIMP to undergo an inelastic transition against another nucleus of mass $\mN$, the total kinetic energy in the center of mass frame must be greater than the inelasticity,
\begin{equation}
\frac{1}{2}\mu w(r)^2 > \exE  \quad \quad \mu = \frac{\mN + \mX}{\mN\mX}
\end{equation}
This condition is more easily satisfied for heavier nuclei as we show in Fig.~\ref{fig:KEvsM} with a plot of ${\rm K.E.}$ for the relevant elements in the Sun as a function of the mass enclosed in a given shell (plotting against the radius is more intuitive, however, it can be misleading since the mass density is not uniform). We took the velocity at infinity to be the most probable speed in a Maxwell-Boltzmann distribution of velocities with a Sun rotation velocity of $220~{\rm km/s}$\footnote{This fixed choice of velocity was made only with regard to Fig. \ref{fig:KEvsM}. In the rest of the paper we used the full Maxwell-Boltzmann distribution for the velocity at infinity as explained in the appendix.} . We see that iron is an effective scatterer throughout the Sun, but oxygen is only useful in the inner 50\% of the Sun's mass and helium is altogether useless as the kinetic energy is never sufficient to overcome the inelastic barrier. Indeed, as we shall see in section \ref{sec:results}, scattering off of iron yields the largest capture rate. This is in contradistinction with the elastic case, where iron places only third behind oxygen and helium because of the form-factor suppression discussed below \cite{Gould:1987ir}. To obtain the results presented in section \ref{sec:results} we included all the different elements persent in the Sun with their proper densities~\cite{Bahcall:2000nu}.  The heavier elements were taken from~\cite{Grevesse:1998bj} and their radial profile was assumed to follow the mass profile of the sun. With the exception of hydrogen and helium, all the different elements included allow for scattering somewhere in the Sun and contribute to the capture rate. 

The inelastic barrier in fact enhances the capture rate in those cases where a scattering is kinematically accessible. In order to be gravitationally captured the WIMP must lose a certain amount of kinetic energy. In the elastic case, the entire energy must be transferred to the nucleus against which the WIMP scatters. In the case of iDM, however, a significant fraction of this energy is lost to the excitation. When the excited state relaxes back to the ground state, it can only do so through the emission of very light states with no significant recoil. The net result is that the inelastic transition allows the WIMP to lose energy efficiently and hence to further enhance the capture rate.   

The transfer of a large fraction of the kinetic energy to an excited state instead of to the nucleus further contributes to the rate by softening the momentum transfer and taming the reduction usually associated with the nuclear form-factor. This is particularly important in the case of iron where the form-factor suppression in the elastic case is very significant and can result in an order of magnitude reduction in rate for $\mX \sim 100\GeV$.  More specifically consider a form-factor $\exp(-Q/Q_0)$ with $Q_0 = 82\keV$ for iron (see appendix \ref{app:caprate} for details) and where $Q$ is the energy transfer. If $Q$ is now smaller by about $\exE \sim 100\keV$ this results in a rate increase by a factor of $\exp(\exE/Q_0) = 3.4$. 

The above considerations lead us to an important conclusion. While one might have naively expected that inelasticity will result in a substantial reduction in the capture rate this is in fact not the case! Iron (which in the elastic case would have dominated capture if not for the form-factor suppression) is kinematically accessible everywhere in the Sun and in fact enjoys an enhancement in its associated capture rate because of the reduced energy transfer and the related softening of form-factor suppression! This enhancement is by and large sufficient to compensate for the loss of helium (and partly oxygen) as a scatterer.

\section{Results}
\label{sec:results}

In this section we present the numerical results for the capture rate of WIMPs in the Sun. Throughout we will assume a WIMP-nucleon cross-section of $\sigma_n=10^{-40}\cm^2$, and note that the resulting capture rate scales linearly with the cross-section\footnote{$\sigma_n=10^{-40}\cm^2$ is the typical WIMP-nucleon cross-section used in iDM models for fitting the DAMA results, see for example Ref.~\cite{Chang:2008gd}. It is in fact somewhat of a lower bound since the needed cross-sections are almost always higher except when $\mX \sim 100\GeV$ and $\exE \lesssim 50\keV$~\cite{Cui:2009xq}.}. Using the DarkSUSY software package version 5.0 \cite{Gondolo:2004sc} we also give the associated muon yield on the Earth from different annihilation channels of WIMPs in the Sun. Many annihilation channels are in fact already constrained by present bounds on the neutrino flux from the Sun as we discuss below. Finally, we discuss the reach of future neutrino telescope experiments.  

\begin{figure}[h]
\begin{center}
\includegraphics[scale=.62]{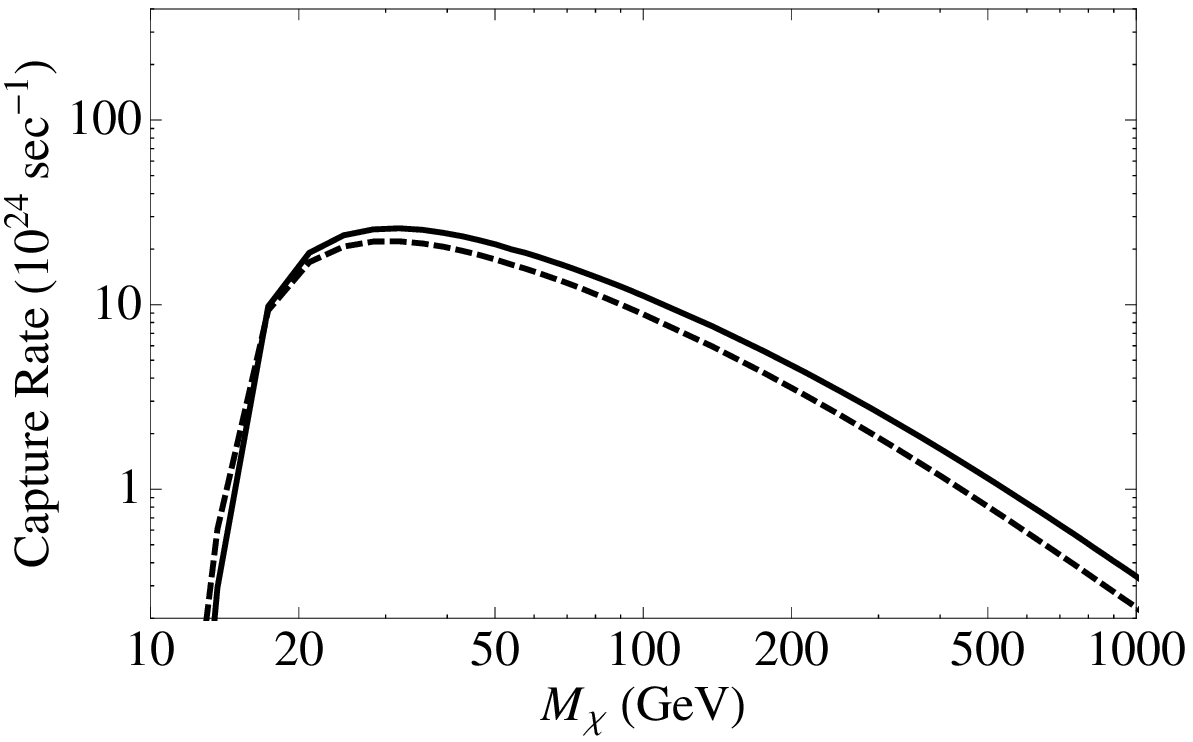}
\includegraphics[scale=.52]{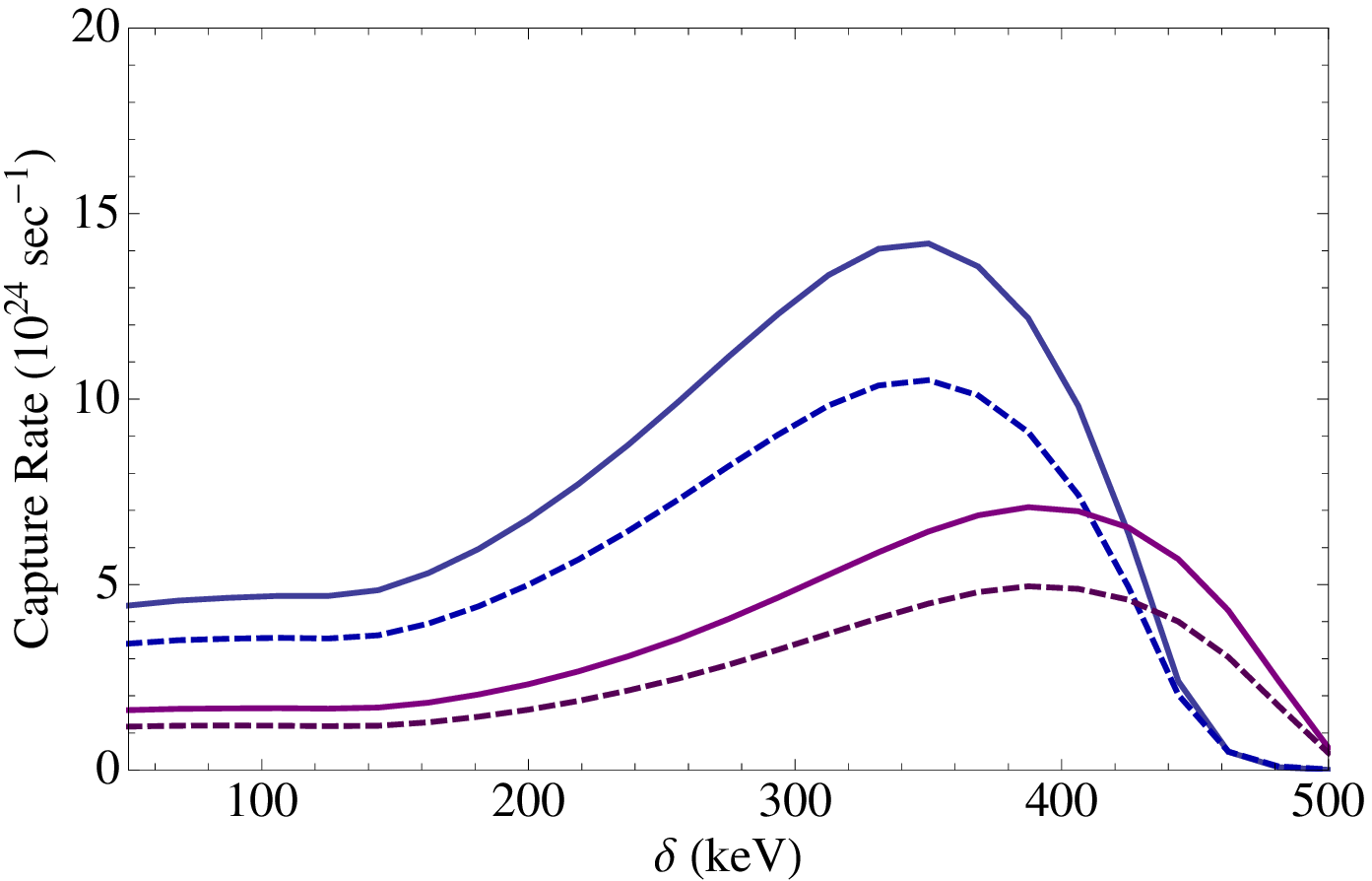}
\end{center}
\caption{On the left pane we plot the capture rate against the WIMP mass. The solid curve correspond to an inelastic model with $\exE =125\keV$ and $v_\odot = 220~{\rm km/s}$ whereas the dotted curve to $v_\odot = 254~{\rm km/s}$. In both cases we take $\sigma_n=10^{-40}\cm^2$. On the right pane we depict the growth of the capture rate as a function of the inelasticity for $v_\odot = 220{\rm km/s}$ (solid) and $v_\odot = 254{\rm km/s}$ (dashed). The upper two curves correspond to $\mX = 200\GeV$ (blue) and the lower two curves to $\mX = 400\GeV$ (purple).}
\label{fig:CaptureRate}
\end{figure}

In Fig. \ref{fig:CaptureRate} we plot the capture rate in the Sun as a function of the WIMP's mass for two different choices of the most probable WIMP velocity in the halo. 
In the same figure we also depict the capture rate's dependence on the excitation energy, $\exE$. As $\exE$ increases, the exponential suppression due to the nuclear form factor is curbed since less momentum is transfered. This results in a rapid rise of the capture rate as illustrated in Eq. \ref{eqn:apxDiffCRFF} in appendix \ref{app:caprate}. However, when $\exE$ becomes too large the capture rate diminishes rapidly because fewer shells in the Sun can participate in the capture. 

\begin{figure}[h]
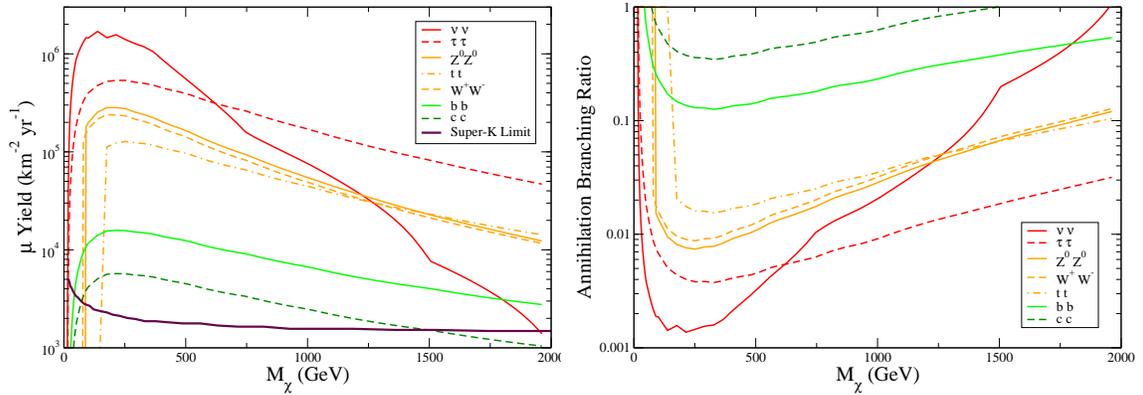

\begin{center}
\includegraphics[scale=.3]{MuonYieldvsMx.eps}
\includegraphics[scale=.3]{BRvsMx.eps}
\end{center}
\caption{On the left is a plot of the muon yield in the inelastic case ($\exE=125\keV$, $\sigma_n =10^{-40}\cm^2$) for different annihilation channels from top to bottom on the left: $\nu_\mu\nu_\mu$, $\tau^+\tau^-$, $Z^0Z^0$, $W^+W^-$, $t\bar{t}$, $b\bar{b}$, and $c\bar{c}$. The area above the thick (violet) curve is excluded by Super-K. On the right pane we plot the corresponding bound on the annihilation branching ratio for the respective channels against the WIMP's mass.}
\label{fig:currentLimits}
\end{figure}

There are existing limits on the flux of muon-neutrinos from WIMP annihilation in the Sun from both underground detectors (BAKSAN~\cite{Suvorova:1999my}, Super-Kamiokande~\cite{Desai:2004pq}, and MACRO~\cite{Ambrosio:1998qj}) as well as dedicated neutrino telescopes (AMANDA~\cite{Ackermann:2005fr}, BAIKAL~\cite{Aynutdinov:2005sc}). The strongest bounds acually come from the Super-K results which we show in Fig. \ref{fig:currentLimits} alongside the muon yield for several annihilation channels plotted against the WIMP's mass. These channels are excluded by several orders of magnitude. On the right of Fig. \ref{fig:currentLimits}, we plot the corresponding limits on the branching ratios of the different annihilation channels. The bounds on direct annihilation into neutrinos deteriorate at higher WIMP mass because the more energetic neutrinos are further attenuated by matter in the Sun. 

\begin{figure}[h]
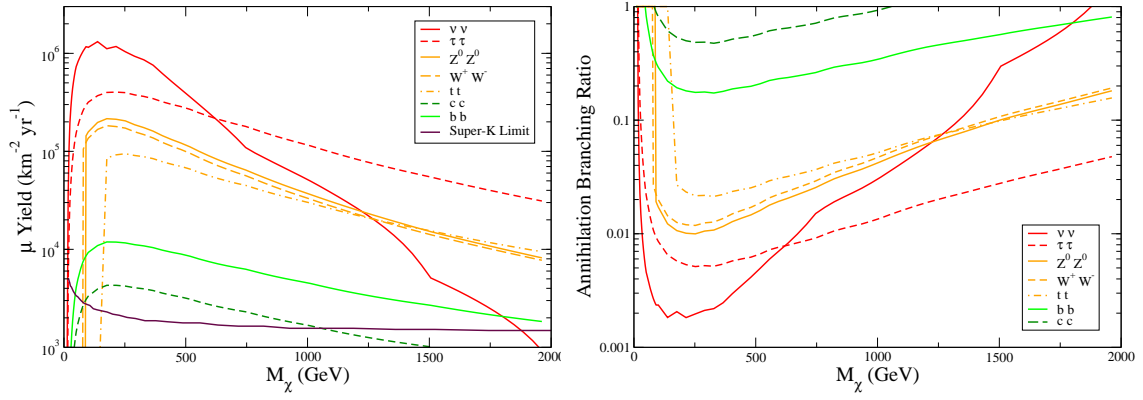

\begin{center}
\includegraphics[scale=.3]{MuonYieldvsMx_v.eps}
\includegraphics[scale=.3]{BRvsMx_v.eps}
\end{center}
\caption{Same as Fig. \ref{fig:currentLimits}, but with $v_\odot = 254~{\rm km/s}$.}
\label{fig:currentLimitsv}
\end{figure}

\begin{figure}[h]
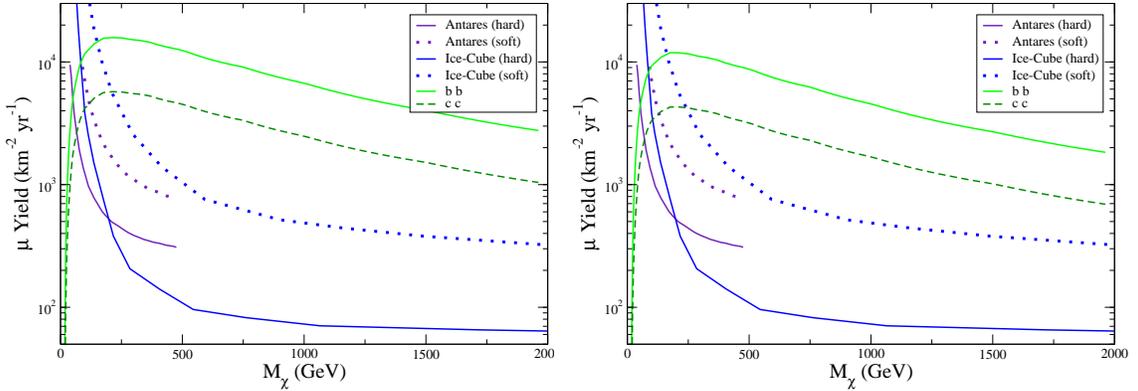

\begin{center}
\includegraphics[scale=.3]{ReachvsMx.eps}
\includegraphics[scale=.3]{ReachvsMx_v.eps}
\end{center}
\caption{The bottom (top) solid curve shows the muon yield from WIMP annihilation in the Sun into $c\bar{c}$ ($b\bar{b}$) which result in a soft muon spectrum for $v_\odot = 220~{\rm km/s}$ ( $v_\odot = 254~{\rm km/s}$) on the left (right) pane. Antares' reach is shown for hard (soft) spectrum in the violet solide (dotted) curve. IceCube's reach is shown with the blue solid (dotted) curve for hard (soft) spectrum. } 
\label{fig:futureLimits}
\end{figure}

Future neutrino telescopes (Antares~\cite{Aslanides:1999vq} and IceCube~\cite{Achterberg:2006md}) are expected to have larger exposures and may provide even stronger constraints on WIMPs annihilation in the Sun. In Fig. \ref{fig:futureLimits} we show the expected reach for both hard and soft spectra together with the expected yield from the least constrained annihilation channels ($c\bar{c}$ and $b\bar{b}$, both resulting in a soft spectrum). We assumed a muon threshold of $5\GeV$ which is appropriate for Antares, but probably too low for IceCube.

Recently, motivated by the positron excess seen in PAMELA, it was suggested that DM may annihilate into two light bosons which are weakly mixed with the SM and so subsequently decay into light particles such as electrons, muons and pions \cite{ArkaniHamed:2008qn}. As mentioned above, such channels cannot  be probed by neutrino telescopes since those final states will stop in the Sun before decaying to neutrinos. However, gauge invariance requires these light bosons to also mix with the $Z^0$ and so have a small coupling to neutrinos as well. In the case of kinetic mixing the coupling to neutrinos compared to the coupling to charge is suppressed by $m^2/M_Z^2$, where $m$ is the mass of the dark vector-boson. Considering the enormous yield from direct neutrino production this might be observable when $m>3\GeV$. However, in that case, the unsuppressed decay into charm quarks is likely to give a better bound. 
\vspace{7mm}

\section{Conclusions}
\label{sec:conclusions}

The iDM scenario offers the exciting possibility of extra structure in the DM sector which explains the discrepancy between the DAMA results and limits coming from other direct detection experiments. It may also be responsible for some unexplained background events in those detectors employing heavier targets (XENON \cite{Angle:2007uj} and CRESST \cite{Angloher:2008jj}). In this paper we showed that such a mechanism will result in a very large capture rate of WIMPs in the Sun. Current bounds on the neutrino flux from Super-Kamionkande strongly constrain the annihilation channels of such WIMPs into SM particles.  Assuming a WIMP-nucleon cross-section of $\sigma_n = 10^{-40}\cm^2$, it requires the branching ratio of annihilations into $W^+W^-$, $Z^0Z^0$, $\tau^+\tau^-$, $t\bar{t}$ and neutrinos to be $< 1\%$. Annihilations into $b\bar{b}$ and $c\bar{c}$, which result in a softer spectrum, are less constrained. Future limits from Antares or IceCube may reduce these branching ratios by an additional order of magnitude.

It is important to realize that these results are likely only robust to within $\sim 50\%$ due to the uncertainties in some of the input parameters (such as WIMP velocity distribution, iron distribution in the Sun, precise form-factors and etc.). Therefore, while the harder annihilation channels are strongly constrained, the softer ones are less so. Finally, it was recently shown in Ref. \cite{Bruch:2009rp} that a dark disc (as compared to the spherical halo assumed in this work) would enhance the capture rate by up to an order of magnitude. If this effect is taken into account, even the softer annihilation channels are strongly constrained. 

Interestingly enough, the recent framework suggested in Ref. \cite{ArkaniHamed:2008qn} unabashedly escapes these bounds. Motivated by the recent results from PAMELA, the DM in this picture annihilates predominantly into light leptons or pions through intermediate light bosons. Such light charged particles stop in the Sun before they decay and do not result in any observable neutrinos. Nevertheless, the bounds discussed in this paper restrict the coupling of DM to the SM. Aside from restricting the possible annihilation channels, it also has a direct impact on the possible mechanisms for the mediation of supersymmetry breaking to the extended DM sector.   

\textbf{Note Added:} We would like to thank the authors of Ref. \cite{Menon:2009qj} for the useful communication regarding their results which are in qualitative agreement with ours. While they also consider the inelastic capture of WIMPs in the Sun, some of their input parameters are different and serve to emphasize the inherent uncertainties in the computation when compared with our results.  
  
\textbf{Acknowledgments}: We would like to thank N. Arkani-Hamed, D. Lynden-Bell, A. Pierce, and N. Weiner for very useful discussions. I. Y. would like to thank the CCPP for their hospitality while this work was completed. L.-T. W. and I. Y. are supported by the NSF grant PHY-0756966 and the DOE grant DE-FG02-90ER40542.
\appendix

\renewcommand{\theequation}{A-\arabic{equation}}
\setcounter{equation}{0}

\section{Capture Rate Computation}
\label{app:caprate}

In this appendix we summarize the steps, approximations, and figures used to compute the capture rate of inelastic DM in the Sun. We also provide an analytic approximation which help elucidate some of the features present in the numerical results. 

We consider a WIMP with velocity $u$ at infinity which is scattered in a region with escape velocity $v$. Hence, its total velocity in that shell is $w^2 = u^2 + v^2$. In order to be captured, it must scatter down below $v$. The capture rate per unit shell in the Sun is given by \cite{Gould:1987ir},
\begin{equation}
\frac{dC}{dV} = \int du \frac{f(u)}{u} w  \Omega(w)
\end{equation}
where $f(u)$ is the WIMP's velocity distribution at infinity. $\Omega(w)$ is the rate per unit time at which a WIMP of velocity $w$ scatter to a velocity less than $v$. It is given by,
\begin{equation}
\label{eqn:Omegaw}
\Omega(w) = \left(n \sigma w\right) \frac{Q_{max} - Q_{cap}}{Q_{max} - Q_{min}}
\end{equation}
where the first factor, $n \sigma w$ is just the rate of scattering, and the second factor embodies the probability of capture. We now address each component separately.

The inelastic non-relatvisitic cross-section is given by,
\begin{equation}
\label{eqn:inelcs}
\sigma = \sqrt{1-\frac{\delta}{\mu w^2/2}} \left(\frac{\mu^2}{\mu_{ne}^2} \right) \left(\frac{f_p^2Z^2 + f_n(A-Z)^2}{f_p^2} \right) \sigma_n
\end{equation}
with $\mu_{ne}$ being the WIMP-\textit{nucleon} reduced mass, and $\sigma_n$ is the WIMP-\textit{nucleon} elastic cross-section. $f_{p(n)}$ determines the relative contribution of protons (neutrons) and we normalize our results to $f_p = f_n=1$. We choose a fiducial value of $\sigma_n = 10^{-40}\cm^2$. The final rate is easily scalable with these quantities. 

The second component in Eq.~(\ref{eqn:Omegaw}) contains $Q_{max}$ ($Q_{min}$) which is the maximum (minimum) energy transfer possible in the scattering process. $Q_{cap}$ is the minimal energy transfer needed for capture,
\begin{eqnarray}
Q_{max} &=& \frac{1}{2}\mX w^2 \left(1 - \frac{\mu^2}{\mN^2}\left(1-\frac{\mN}{\mX}\sqrt{1-\frac{\exE}{\mu w^2/2}} \right)^2 \right) - \exE \\
Q_{min} &=& \frac{1}{2}\mX w^2 \left(1 - \frac{\mu^2}{\mN^2}\left(1+\frac{\mN}{\mX}\sqrt{1-\frac{\exE}{\mu w^2/2}} \right)^2 \right) - \exE \\
Q_{cap} &=& \frac{1}{2}\mX \left(w^2-v^2\right)  - \exE 
\end{eqnarray}
where $\mu$ is the WIMP-\textsl{nucleus} reduced mass, and $\exE$ is the inelastic energy gap. Including nuclear form factors, the formula for the capture rate per unit time gets modified to,
\begin{equation}
\Omega(w) = \frac{n \sigma w}{Q_{max} - Q_{min}}\int_{Q_{cap}}^{Q_{max}} e^{-Q/Q_0}
\end{equation}
where,
\begin{equation}
Q_0 = \frac{3}{2\mN R^2} \quad \quad R = 10^{-13}~\left(0.91\left(\frac{\mN}{\GeV}\right) + 0.3 \right)\cm
\end{equation}

We assume that the WIMP's velocity distribution in the halo is given by a Maxwell-Boltzmann distribution as seen by an observer moving with velocity $\vsun$
\begin{equation}
 f(x) dx = \frac{\rhoX}{\mX} \frac{4}{\sqrt{\pi}} x^2 e^{-x^2} e^{-\eta^2}\frac{\sinh (2 x\eta)}{2x\eta}dx
\end{equation}
where $\rho=0.3\GeV \cm^{-3}$ is the local mass density and the dimensionless quantities $x$ and $\eta$ are given by,
\begin{eqnarray}
x^2 &\equiv& \frac{u^2}{v_0^2} \\
\eta^2 &\equiv& \frac{\vsun^2}{v_0^2}
\end{eqnarray}
with $v_0 = 220{\rm km/s}$. 

It is possible to obtain approximate analytic results for the capture rate as follows. Because of the phase-space suppression in Eq. (\ref{eqn:inelcs}) the scattering has larger support when $\exE < \mu w^2/2$. Hence we can Taylor expand the radical in the expression for the momentum transfer and get,
\begin{equation}
Q_{max} = \frac{1}{2}\mX v_0^2 \frac{\beta_+}{\beta_-}\left( A^2 + \beta_- x^2 - \left(\frac{\mX+\mN}{\mX-\mN}\right) \frac{\exE}{\mX v_0^2/2}\right) - \exE
\end{equation}
with $\beta_\pm \equiv 4\mX\mN/(\mX\pm\mN)^2$, $A^2 \equiv v^2\beta_-/v_0^2$ (our notation here follows that of Gould~\cite{Gould:1987ir}). In this case, ignoring form-factor for the moment, the integral over the velocity distribution can be done exactly and the results are identical to those found in Ref. \cite{Gould:1987ir} with the identification,
\begin{equation}
\label{eqn:Aidn}
A^2 \rightarrow A^2 - \left(\frac{\mX+\mN}{\mX-\mN}\right) \frac{\exE}{\mX v_0^2/2}
\end{equation}
This tends to reduce the capture since it diminishes the maximum energy transfer and hence the probability for capture. Notice that the additive term involving $\exE$ in the energy transfer drops out since in the absence of form-factors the capture rate involves only differences in energies. 

This term becomes significant when including the form-factor as it serves to \textit{increase} the capture rate. Using Eq. (\ref{eqn:Aidn}) it is possible to obtain an approximate analytical expression in this case as well. Regardless, the most important feature is that the capture rate grows exponentially with $\exE$,
\begin{eqnarray}
\label{eqn:apxDiffCRFF}
\frac{dC}{dV} = \frac{2 \sigma n \rhoX v_0}{\sqrt{\pi}\mX} e^{\exE/Q_0}\left( \ldots \right)
\end{eqnarray}
where the ellipsis denote the usual expression involving the form-factor corrected by $\exE$. This behavior reflects the fact that as the inelastic threshold increases, less energy is transferred to the nucleus and the form-factor suppression is correspondingly smaller!
Therefore, in the case of inelastic transitions, the reduction due to form factor is not as severe. In particular it allows iron to assume the role of the prime contributer to capture, a role it would have played in the elastic case as well if not for the form-factor suppression.  

All together, inelasticity enters in two significant ways. First, it tends to reduce the capture rate because of the diminishing phase-space. Second, since less energy is transferred to the nucleus, the reaction is more coherent, form-factor effects are diminished, and the capture rate enjoys an increase as compared with the elastic case. For iron in particular,  the second effect is much more important as discussed in the text above.

\renewcommand{\theequation}{B-\arabic{equation}}
\setcounter{equation}{0}

\section{Density Computation}
\label{app:density}

In order to compute the density of the WIMPs accumulated in the Sun we approximated the Sun's potential with an analytic formula which allows for an exact solution for the orbits \cite{Henon1, Henon2}. It interpolates between an harmonic potential close to the core and an inverse fall-off at long distances, 
\begin{equation}
U(r) = \frac{G M_{\odot} \mX}{2 bR_{\odot}}\left( 1- \frac{2b}{b+ \sqrt{b^2+r^2}}\right)
\end{equation} 
(in here the potential energy is computed with respect to the origin where it is zero rather than at infinity). A fit to the Sun's actual potential \cite{Bahcall:2000nu}, gives $b = 0.0884 R_{\odot}$ and yields an approximation which is roughly $10\%-20\%$ accurate. The orbits are found by first expressing the potential $U$ in terms of $r$, which by an adequate choice of units can be written simply as,
\begin{equation}
r^2 = \frac{U}{\left(1-U\right)^2}
\end{equation}
The dependence of the azimuth and the time variable on the potential (and hence on $r$) can be found by usage of the constants of motion and a straightforward integration. The interested reader can find the details of the solution in Ref. \cite{Henon2}.  

\begin{figure}[h]
\begin{center}
\includegraphics[scale=.62]{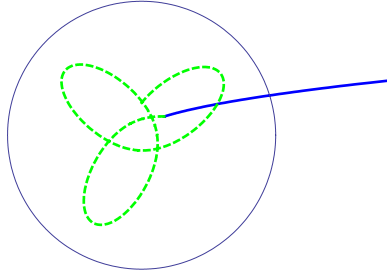}
\end{center}
\caption{A typical event in which a WIMP approaches the Sun, collides with an Iron nuclei and gets captured in a bounded rosette-like orbit. The solid-blue line is the approach path from infinity to the point of collision. The dashed-green line is the orbit after collision which is normally contained within the Sun that is delineated by the circle.}
\label{fig:orbit}
\end{figure}

To find the final density, we incorporated the exact solution for the orbits into a Monte-Carlo simulation. We let the WIMP's start at infinity with an impact parameter which is uniformly distributed and a shifted Maxwel-Boltzmann velocity distribution with the rotational speed of the local standard of rest $v_0 = 220~{\rm km/s}$ and $v_{\odot} = v_0$. The WIMP is then allowed to scatter at any radius in the Sun, assuming its orbit reaches that point and its kinetic energy is sufficiently large to overcome the inelasticity. We take the probability of scattering as a function of radius to be proportional to the local density divided by the WIMP's velocity at that radius. The scattering is assumed to be isotropic in the center of mass frame, and the kinetic energy of the outgoing WIMP is restricted to be below the capture rate. After the collision, the kinetic energy of the outgoing WIMP and its angle with respect to its original direction of motion are used to determine its new orbit. Fig. \ref{fig:orbit} depicts a typical approach and capture of a WIMP by the Sun. If any further collisions are energetically possible then they are executed following similar steps as described above. Once no further collisions are possible the orbit is recorded and stored. 

The density is computed by simulating 10,000 capture events and then randomly sampling the radii of the resulting orbits. It is important to sample uniformly in time rather than radius because the particle actually spends most of its time close to the apogee. 

\bibliographystyle{JHEP}
\bibliography{suncapture}
\end{document}